# THE BLACK SEA WAVE ENERGY: THE PRESENT STATE AND THE TWENTIETH CENTURY CHANGES

Vasko Galabov

**Abstract**

In this paper we present a study of the present state of the Black Sea wave energy. The studies of other authors are based on the use of input data from atmospheric reanalysis or a downscaling of such reanalysis. Instead of reanalysis data, we use input data from the operational limited area numerical weather prediction model ALADIN. We showed that the estimations of the Black Sea wave energy based on reanalyses deviate significantly from the real potential. We showed also that the highest values of the mean annual wave power flux is between 4.5 and 5.0 kW/m$^2$ and the near shore areas with the highest wave energy potential are the southernmost Bulgarian coast and the coast of Turkey north of Istanbul. While we showed that the wind data from the reanalysis are not useful for the estimation of the actual wave energy potential, we claimed that the reanalysis data is useful to study the long term changes of the wave energy of the Black Sea. We used the 10m winds from the recent ERA-20C reanalysis, which covers the period 1901-2010. We performed a 110 years hindcast with these winds using the SWAN wave model. The results for the area with the highest mean annual wave energy showed that there was an increase during the first half of the XX century followed by a small decrease and again a period with elevated wave energy during the seventies. After 1980 there is a steady decrease of the Black Sea wave energy.

**Introduction.** During the last decade there is a growing interest in the field of marine energy as a renewable energy resource. The Black Sea is tideless and with regards to the energy of the wind waves it is considered a low energy environment, but nevertheless there are studies of the wave energy potential of the Black Sea. The wave energy in the Black Sea as a potential renewable resource was first studied by E. Rusu [1]- in this study the spatial patterns of the wave energy for some typical cases were assessed. Akpinar and Komurcu [2] estimated the mean annual wave power flux (hereafter denoted MAWPF) in the Black Sea by numerical hindcast based on input wind data with 6 hour temporal resolution from ERA- Interim [3] reanalysis. Their estimation shows that the MAWPF of the Black Sea is below 3 kW/m and the mean annual significant wave height (significant wave height is hereafter denoted SWH) of the Black Sea is below 0.8 m. This value of the mean annual SWH is in agreement with the estimation of Stanev and Kandilarov [4] based on numerical simulation using wind data from dynamical downscaling of global reanalysis. Meanwhile Aydogan et al [5] published a numerical hindcast based on wind data from ECMWF product (unspecified) also with 6 hour temporal resolution. They estimated that the MAWPF reach a maximum value above 7 kW/m for the Western Black Sea and values above 6 kW/m for some nearshore locations. The next study of the Black Sea wave energy potential was published by Valchev et al [6]. They used a dynamical downscaling of a global reanalysis data to force the wave simulation and also their hindcast is based on a longer period of 60 years. Their estimation of the MAWPF is slightly higher than the estimation of Akpinar and Komurcu- up to 3.5 kW/m and they argued that the higher value is due to the inclusion of the period before 1980 which is well known with the higher number and intensity of the storms in the Black Sea [7-8]. L. Rusu and Onea [9] used directly the wave parameters in ERA- Interim reanalysis in order to estimate the MAWPF at some locations close to the Western Black Sea coast and their conclusions is that it can reach values above 4 kW/m for the North Western Black Sea shelf. As a conclusion from this short overview of the studies of the Black Sea wave energy potential there is some discrepancy between the estimates. All previous studies are based on reanalysis data (the wind data source in the study of Aydogan et al is not properly specified but it is also not an operational product, because there was no operational model of ECMWF with the declared spatial resolution back in 1996 and most likely a downscaling of reanalysis). The reanalysis data has limitations with regards to the estimations of the actual state of the wave energy potential of the Black Sea- for instance the study of Kara et al [10] showed that ERA reanalysis significantly underestimates the Black Sea winds while NCEP reanalysis [11] overestimates the Black Sea winds and is considered by the authors not useful at all for Black Sea applications due to too high discrepancy with the measured winds and too high level of errors. We confirmed this in the present study. As we argued in our previous work [12] the way to obtain a reliable estimation of the present state of the Black Sea wave power is to use an alternative source of wind data from a high resolution (both in spatial and temporal meaning) limited area model that was tuned up especially for the Black Sea application. The main aim of this study was to do this by using the operational limited area model ALADIN specifically set up at the National Institute of Meteorology and Hydrology (NIMH) of Bulgaria not only for the routine weather prediction inland but also for the Black Sea. Another aim of the present paper is to study the long term changes in the Black Sea wave power and mean annual SWH. The recent global reanalysis ERA 20C [13] covers the period 1900-2010 and may be used to perform a long term hindcasts of the wave parameters, taking into account that in ERA 20C only surface data is used and this way it is more homogeneous in time than the previous reanalyses. Also it is available with a temporal resolution of the data of 3 hours which is important for applications such as wave modelling in semi enclosed seas. We present 110 years hindcast of the MAWPF and mean annual SWH of the Black Sea and discuss the relation of the changes to the global teleconnection indices in order to identify the driving factors for these changes. Finally we should point that the study of the wave energy (with wave power flux as the main indicator) is important not only in the frame of the renewable energy sources research, but equally important as a parameter of the wave climate.

**Data and methods.** In the present study we use the SWAN wave model [14] version 40.91ABC. Detailed explanation of the model application for the Black Sea is available in [1-2,8]. While the present version of the model is 41.01, we decided to use this version instead of the most recent one, due to the change of the wind drag formulation from linear bulk formula to nonlinear that leads to significantly lower wind drag values for wind speed above 20m/s. Because of slight underestimation of the strong winds by the operational model that we use in the present study, this leads in our operational practice to poor performance of SWAN used as a forecasting tool. Note that this problem does not exist when the WAM cycle IV physics is used (hereafter denoted as Janssen) [15] due to the explicit way to obtain the friction velocity in Janssen's theory). SWAN computational grid in this study is regular spherical grid with 1/30° spatial resolution and based on discretisation in 36 directions and 31 frequencies between 0.05 and 1 Hz. The model domain covers the entire Black Sea and the number of computational nodes is 451x211. The interval between the outputs is set to 1 hour. We compare SWAN simulations of the significant wave heights for a very energetic period from 20 January 2012 to 10 February 2012 using satellite altimetry data from JASON 1 and ENVISAT satellites. The reason for that is that it is very important for such kind of studies to ensure that the model performs well not only for average but also for high energy conditions. We test four different combinations of parameterizations of the wave generation and dissipation due to whitecapping- the already mentioned WAM cycle IV parameterization scheme, the WAM cycle III parameterization scheme (hereafter denoted Komen) [16] with two values of δ in the whitecapping parameterization δ=1 (default starting with the version of SWAN 40.91.ABC) and δ=0 (default in the previous versions) and we also test the parameterizations of Westhuysen [17] based on saturation based whitecapping (instead of pulse based in the other two) and alternative version of the generation by wind of Yan (but using bulk formula for the wind drag coefficient). The wind input data for the wave model is from ERA-20C reanalysis for 110 years and also ERA-Interim for reference and NCEP reanalysis II. While CFSR is also an option, when it comes to long term climatic changes it is too short (a bit more than 30 years) to reveal the changes of the wave climate for a long enough period. To estimate the present state of the Black Sea wave power we use the operational limited area atmospheric model ALADIN [18] used routinely to drive the Bulgarian operational wave forecast and storm surge forecast system. We take the initial model otput +0 hours and also the model forecast for +3 hours, +6 hours and +9 hours and then we continue with the initial output of the next model run (the model starts twice daily). The horizontal resolution of the wind data from ALADIN is 0.125°. The hindcast with ALADIN data covers the period 01.06.2011 to 31.05.2015 because the specific setup of ALADIN is homogeneous in time for the selected period and without significant changes in the performance with regards to the winds speeds for the Black Sea. In our recent study [20] we used dynamical downscaling of ERA-Interim data with ALADIN model to reconstruct some notable historical storms. The wave power flux is estimated by the formula $P=0.486*H^2*T_e$ where H is the SWH and $T_e$ is the energy period (available as a standard output parameter in SWAN). In our simulations the year does not start at the beginning of January, but in low energy month – we selected first of June as the beginning of the year to ensure that the entire high energy winter season is within one year and not artificially separated and so for instance year 2012 means the period 01.06.2011 to 31.05.2012.

**Comparison of different wind sources and parameterizations with satellite data.** We compared with satellite altimetry data the simulations by SWAN of the SWH for the specified in the previous section wind sources and parameterizations for a period of 20 days with very high wave energy for the Black Sea. The results are summarized in table 1.

Table 1

Comparison of SWAN model runs with different input winds and different source terms parameterisations with satellite altimetry measurements of SWH from Jason 1 and Envisat satellites. The period is 20.01.2012 to 10.02.2012. RMSE denotes root mean square error, R-correlation coefficient and SI the scatter index.

| Wind data source | source terms parameterization | SWH observation mean | SWH model mean | Bias | RMSE | R | SI |
|---|---|---|---|---|---|---|---|
| ALADIN | Westhuysen | 2.93 | 2.55 | -0.38 | 0.68 | 0.80 | 0.23 |
|  | Janssen |  | 2.57 | -0.36 | 0.63 | 0.81 | 0.21 |
|  | Komen |  | 2.53 | -0.41 | 0.69 | 0.78 | 0.23 |
|  | Komen, δ=1 |  | 2.77 | -0.16 | 0.59 | 0.90 | 0.20 |
| ECMWF analysis | Komen, δ=1 |  | 2.50 | -0.43 | 0.70 | 0.75 | 0.24 |
| ERA-Interim | Komen, δ=1 |  | 2.25 | -0.68 | 0.90 | 0.55 | 0.31 |
| NCEP Reanalysis II | Komen, δ=1 |  | 3.37 | +0.44 | 1.27 | 0.75 | 0.43 |

The bias of the SWH when using the ALADIN wind input data is lowest when we use the Komen option with δ=1 (and also the RMSE and scatter index are the best) and highest (i.e. highest underestimation) when using Komen with δ=0. Because the estimation of the wave power flux depends on the square of SWH, high biases are undesirable because they can easily lead to erroneous results due to multiplication of errors. Therefore we chose Komen, δ=1 in our hindcasts. The simulations based on ECMWF operational analysis and ERA-Interim lead to too high negative biases of SWH (which explains the low wave power of the Black Sea estimated by Akpinar and Komurcu). The NCEP Reanalysis II leads to high overestimation and high scattering, making it inapplicable for such studies in the Black Sea (not to mention also the too coarse resolution of this reanalysis. Obviously the usage of ALADIN wind data outperforms all reanalyses for the SWH and is preferable for wave power estimations also for it high resolution.

**Estimation of the present state of the Black Sea wave power.** The estimation of the Black Sea wave power (MAWPF) and the mean annual SWH for the period 01.06.2011- 31.05.2015 when using ALADIN wind input is presented on fig.1.

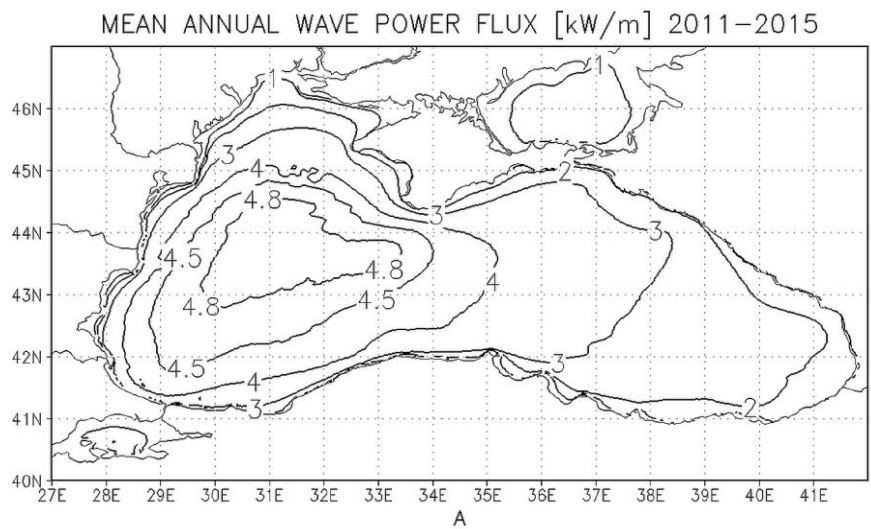

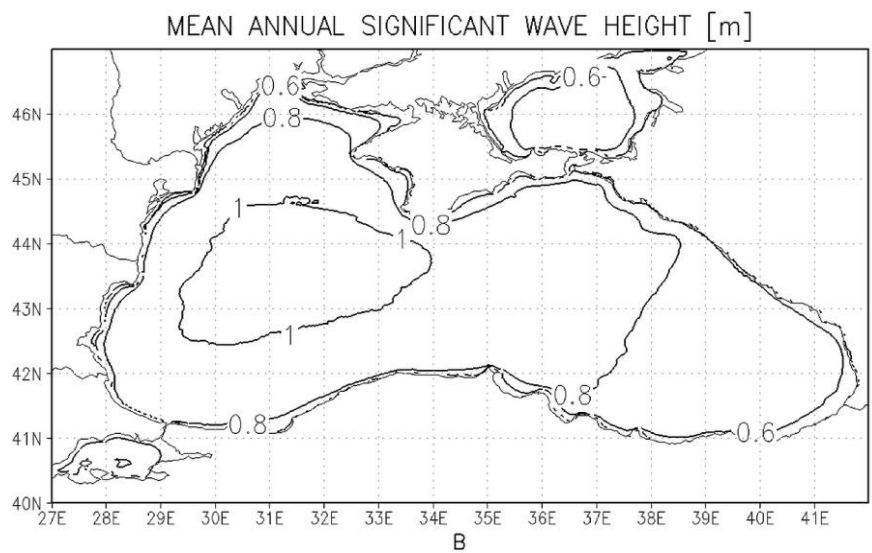

Fig. 1. SWAN hindcast with input wind data from ALADIN atmospheric limited area model for the period June 2011 to May 2015. A) mean annual wave power flux (MAWP). B) mean annual significant wave height.

As it can be seen the MAWPF reach values of about 4.8 kW/m for the recent four years and mean annual SWH- slightly above 1m. The areas with the highest wave power in agreement with other studies are the Turkish coast north of Bosporus and the southern Bulgarian coast but also in some areas in north-western shelf. This is above the estimations based on reanalysis but below the values obtained by Aydogan et al. A possible reason is that in their study the presented validation of their model shows that they are working with positive biases for the SWH and wave period for the most energetic measurements location presented and two positive biases easily lead to significant overestimations and should be avoided in any way.

Table 2

Mean annual wave power flux for some selected points for the period 2012-2015.

| Name of the point | Latitude,° | Longitude,° | Description of the location | Mean annual wave power flux, kW/m |
|---|---|---|---|---|
| Ahtopol | 42.20 | 28.20 | Near Ahtopol- southernmost Bulgarian coast | 4.00 |
| Shabla | 43.60 | 28.90 | Near cape Shabla- close to the border Bulgaria-Romania | 4.01 |
| North Western Shelf | 44.81 | 30.00 | In the North Western Shelf close to Danube delta | 3.88 |
| Crimea | 42.27 | 33.70 | South of Crimea Peninsula | 3.48 |
| Gelendzhik | 44.50 | 37.98 | Close to Gelendzhik, Russia- north-eastern Black Sea | 2.20 |
| Sinop | 42.19 | 35.0 | Close to Sinop- Turkey- central southern coast | 3.73 |
| Batumi | 41.65 | 41.35 | Close to Batumi- south-eastern coast | 1.79 |
| Bosporus | 41.50 | 29.0 | North of Bosporus-south western coast | 4.22 |

In table 2 we present the estimation of the MAWPF for 8 selected locations. All locations are with depth in the model bathymetry between 50 and 100m (so all of them are in deep water with regards to wave modelling) and 10 to 30km from the coastline. As it can be seen, the wave power is above 4 kW/m for the Western Black Sea shelf and lowest for the south eastern Black Sea.

Figure 2 shows the variations of the mean monthly wave power flux for the location Ahtopol. As it can be seen, the winter months are with mean wave power reaching for some months values above 10 kW/m, while the summer months are with low wave power. The highest value is for February 2012 but the year with the highest mean annual value is 2015 (01.06.2014-31.05.2015) with MAWPF for Ahtopol above 5 kW/m. Also when ERA-Interim winds are used, the wave power is significantly underestimated for the months with the highest energy. The estimation shows that the waves with SWH above 1m have a contribution to the MAWPF of 87% and the waves above 2m- 56% share of MAWPF but their frequency is low- the waves are above 1m 29% of the total time of the hindcast and these above 2m just 7% of the time. So the value of the MAWPF is mainly due to contribution of short lasting high energy episodes and very low energy otherwise. With regards to directionality of the wave power, for Ahtopol the energy comes mainly from northeast with relatively low directional spreading, but generally there are no significant differences with the other authors about the directions.

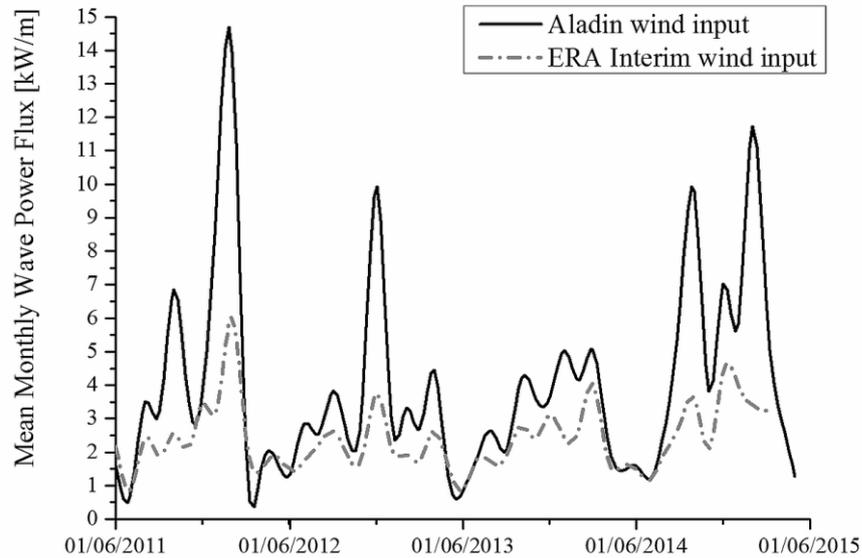

Fig. 2. Variation of the mean annual wave power flux during the period 06.2011-05.2015 at Ahtopol location based on ALADIN and ERA-Interim wind input data.

**Changes of the Black Sea wave energy for the last 110 years.** To evaluate the changes during the last 110 years of the mean annual SWH and the wave energy, we performed a 110 years simulation with SWAN using ERA-20C wind input. The comparison of simulations with ERA-20C and ERA-Interim winds for the period 1979-2010 show that the results with the two reanalyses are consistent with slightly higher values when using ERA-20C due to the higher temporal resolution of the data. While in the table 1 we presented comparison with measurements for the entire Black Sea, we must mention, that the statistics for the Western Black Sea are significantly better for both ALADIN and ERA-Interim (and so the same is expected for ERA-20C) while for the Eastern Black Sea they are significantly poor. Due to that reason we consider the long hindcast conclusions reliable only for the western part and because for the eastern part they may be artefacts of the reanalysis inaccuracy, we will make conclusions only for the Western Black Sea that is more energetic anyway and therefore much more interesting in the frame of the present study.

Figure 3 presents the changes of the wave power for Ahtopol and Shabla locations. A smoothed curve of the changes was obtained by use of low pass Fast

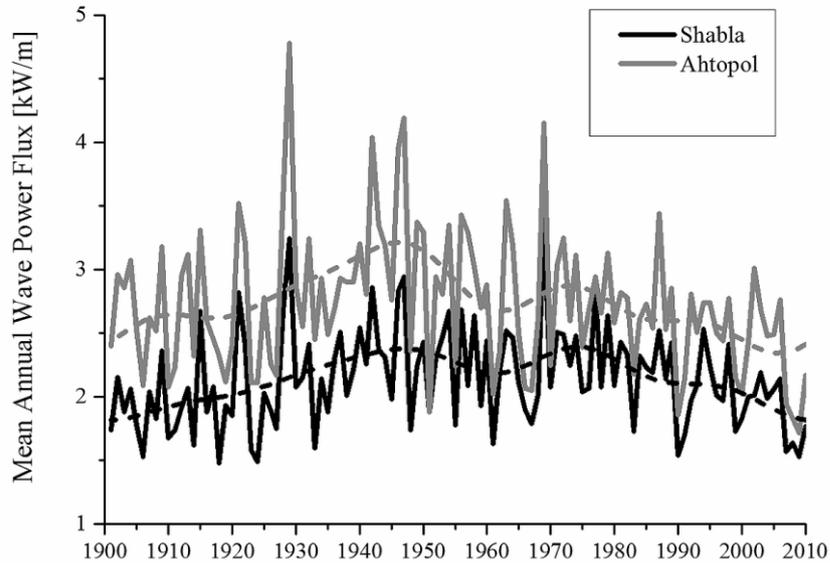

Fig. 3. 110 years hindcast (based on ERA-20C winds) of the mean annual wave power flux at Ahtopol and Shabla locations- annual data and smoothed by low pass FFT filter with 0.05 cut off frequency.

Fourier Filter (FFT) with cut of frequency of 0.05. The form of the curve is qualitatively the same when using moving average with 10 years period of averaging. As it can be seen there is for both locations increase during the first 50 years of the XX century (the positive trend is statistically significant at level of significance 0.01 using Mann-Kendal test). After that there is a decrease with some increase during the seventies and negative trend after that (also statistically significant at level 0.01 by Mann-Kendal test). We found also that the changes are mainly due to the changes of the wave regime of the waves below 4m SWH that follow such trend, while for the waves above 4m SWH the behaviour is different- there are no signs of trends, but rather of a low frequency oscillations such as those found by Polonsky et al [7] for the storms in the Northern Black Sea with a period of 50-60 years. For Crimea there is a steady increase of the wave energy for the entire period until the end of seventies and decrease after that. Even if we argued that the results for the eastern part of the sea are not so reliable, we should mention that the seventies are particularly interesting, because there is a high value of the wave energy for the entire Black Sea. The maximum decadal mean wave power of the Black Sea (the highest value in the entire sea) for the 110 years period increased from 2.7 kW/m (remember that due to the underestimation the absolute values are not important but only the dynamics) to 3.3 kW/m during the decade 1941-1950. Then it decreased to 3 kW/m in the period 1951-1970 and raised again to 3.3 kW/m during the seventies and after that decreased to 2.7 in the last decade (see figure 4). These changes need explanation and the way to do it is to search for the correlations with the global atmospheric indexes (teleconnections) and other climatic indexes.

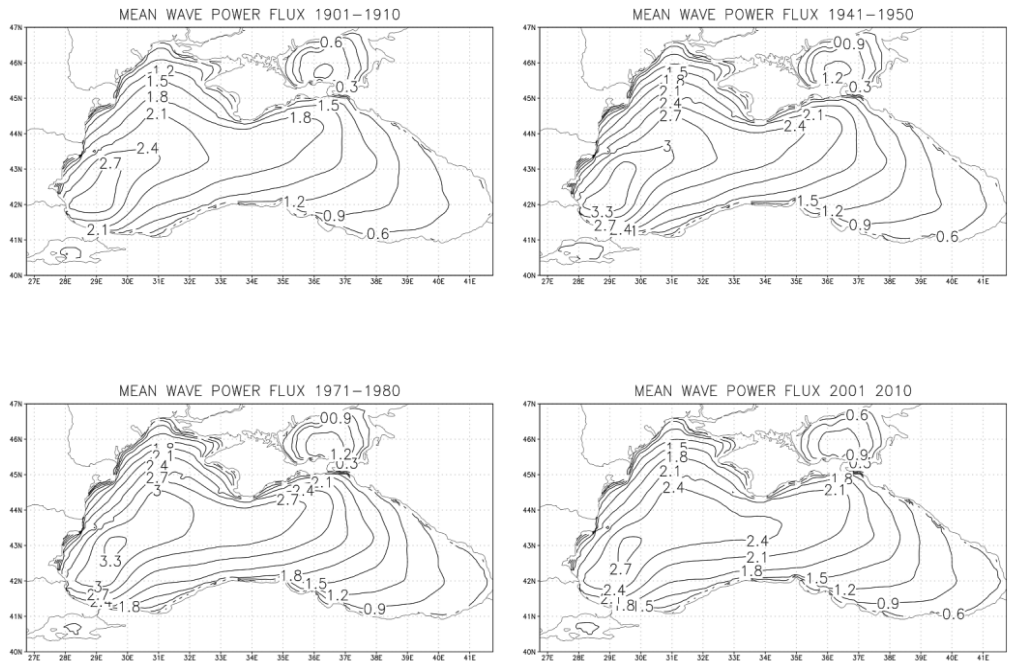

Fig. 4. Mean decadal wave power flux changes- upper left: 1901-1910; upper right: 1941-1950; lower left: 1971-1980; lower right: 2001-2010

**Links of the changes to global teleconnections.** We studied the correlations of the mean monthly and annual wave power to the global teleconnections such as North Atlantic Oscilation (NAO), East Atlantic/West Russia (EA/WR) pattern, Southern Oscilation Index (SOI), Arctic Oscilation, Scandinavia index and others and also the patterns in the ocean circulation Atlantic Multidecadal Oscilation (AMO) and Pacific Decadal Oscilation (PDO). For the Northern Black Sea Polonsky found that the main factor driving the low frequency oscillations of the storminess there is the interplay of AMO and PDO. Our study confirms this finding also for the wave power. Multivariate regression shows that the correlations of the wave power for Crimea with AMO and PDO are significant and the combined AMO+PDO oscillation correlates with the wave power changes there with correlation coefficient R=0.61. For the Western Black Sea shelf this however is not the case. The correlation with AMO and PDO is not significant and is below 0.1 and obviously the processes that determine the wave power changes and so the average wave climate (not the storminess climate that needs further studies) is affected by different reasons. Only the correlation with NAO and EA/WR was found to be significant at p-value of 0.05 using Pearson test and there is a higher correlation with EA/WR than the correlation with NAO- for Ahtopol the correlation with EA/WR is 0.35 and with NAO is 0.27. The multivariate regression analysis shows that these two teleconnections with some contributions of the others increase the correlation with the teleconnections to 0.49 and obviously significant part of the variations remains unexplained by the global processes. For the north western shelf (Shabla location) there is a higher correlation with NAO- 0.33 and 0.25 with EA/WR. In general the correlation with NAO decrease from north to the south at the Western Black Sea.

Surkova et al [20] used a future climate projection of the Black Sea storminess and concluded that the storminess of the Black Sea is expected to decrease and we may speculate that this also means a further decrease of the Western Black Sea wave energy as well, but taking into account how big part of it variations remain unexplained, this remains only a speculation.

**Conclusions**. The present state of the Black Sea wave power was estimated based on the period 2012-2015, using wind data from the limited area atmospheric model ALADIN. It was found that the mean annual wave energy flux reach 4.8 kW/m for the South Western Black Sea and above 4 kW/m for the western shelf. 110 years wave hindcast was performed to evaluate the changes in the wave power and it was found that the wave power increased during the first half of the XX century for the western part of the sea (where it is highest) and decreased after the seventies. The study of the influence of the teleconnections showed that the changes in the wave power at the western shelf are driven by other factors (mainly linked with NAO and EA/WR) than the northern and eastern part of the sea, where it is linked with AMO and PDO and highest when they are both negative. As for the applicability of the wave energy as a renewable energy resource the conclusions taking into account the negative trend and climate projections are hardly optimistic and it may have some applications only in a combined wind-wave energy converters.

**Acknowledgements.** This study has been prepared with the financial support of the European Social Fund through Project *BG051PO001-3.3.06-0063*. NIMH - BAS is solely responsible for the content of this document, and under no circumstances can be considered as an official position of the EU or the Ministry of Education and Science.

*National Institute of Meteorology and Hydrology*
*Bulgarian Academy of Sciences*
*66, Tsarigradsko Shose*
*1784 Sofia, Bulgaria*
*e-mail:* vasko.galabov@meteo.bg